\documentclass[10pt]{article}
\usepackage{longtable}
\usepackage{amsmath}
\usepackage{enumerate}
\usepackage{array}
\usepackage{subfig}
\usepackage{graphicx}
\usepackage{float}
\usepackage{amsfonts}
\usepackage{epstopdf}
\usepackage[cm]{fullpage}
\usepackage{multicol}
\usepackage{authblk}
\usepackage{url}
\usepackage{cite}
\usepackage{booktabs}
\usepackage{titlesec}
\usepackage[titletoc,title]{appendix}
\usepackage{soul}

\titleformat{\section}
  {\normalfont\sffamily\Large}
  {\thesection}{0.5em}{}
  
\titleformat{\subsection}
  {\normalfont\sffamily\large}
  {\thesubsection}{0.5em}{}

\titleformat{\subsubsection}
  {\normalfont\sffamily\normalsize}
  {\thesubsubsection}{0.5em}{}

\titlespacing*\section{0pt}{20pt plus 4pt minus 2pt}{5pt plus 2pt minus 2pt}
\titlespacing*\subsection{0pt}{16pt plus 4pt minus 2pt}{4pt plus 1pt minus 1pt}
\titlespacing*\subsubsection{0pt}{12pt plus 4pt minus 2pt}{2pt plus 0pt minus 0pt}

\title{\vspace{-1.5cm}\sffamily The reachability of contagion in temporal contact networks: how disease latency can exploit the rhythm of human behavior}
\date{}
\author{Ewan Colman\footnote{ec975@georgetown.edu}}
\author{Kristen Spies}
\author{Shweta Bansal}
\affil{\small{Department of Biology, Georgetown University, Washington, DC 20057, U.S.A}}

\begin{document}
\maketitle
\vspace{-1cm}
\begin{abstract}
\normalsize
\noindent
\textbf{Background:} The symptoms of many infectious diseases influence their host to withdraw from social activity limiting their own potential to spread. Successful transmission therefore requires the onset of infectiousness to coincide with a time when its host is socially active. Since social activity and infectiousness are both temporal phenomena, we hypothesize that diseases are most pervasive when these two processes are synchronized.
\vspace{2mm}

\noindent
\textbf{Methods:} We consider disease dynamics that incorporate a behavioral response that effectively shortens the infectious period of the disease. We apply this model to data collected from face-to-face social interactions and look specifically at how the duration of the latent period effects the reachability of the disease. We then simulate the spread of the model disease on the network to test the robustness of our results.
\vspace{2mm}

\noindent
 \textbf{Results:} Diseases with latent periods that synchronize with the temporal social behavior of people, i.e. latent periods of $24$ hours or $7$ days, correspond to peaks in the number of individuals who are potentially at risk of becoming infected. The effect of this synchronization is present for a range of disease models with realistic parameters.
 \vspace{2mm}
 
 \noindent
 \textbf{Conclusions:} The relationship between the latent period of an infectious disease and its pervasiveness is non-linear and depends strongly on the social context in which the disease is spreading.   

\end{abstract}
\vspace{0.5cm}
\begin{multicols}{2}

\section{Introduction}
Identifying the fundamental causes of epidemic outbreak remains a major challenge that crosses several scientific disciplines. While advances in infectious disease virology remain crucial to preventing epidemics, many contributions have also come from the study of social behavior and the structure of contact networks \cite{Funkrsif20100142,RevModPhys.87.925}. Recent work in this area has begun to address the effect of dynamic contact networks on contagion processes \cite{holme2015modern,2016temporal}, however, the complex interplay that emerges from coupling the dynamics of human social behavior with the life-cycle of an infectious disease is yet to be fully explored. 

After entering a human host, infections typically experience a \emph{latent period} for which the host is unable to infect others. This is then followed by the \emph{infectious period} during which disease transmission is possible \cite{anderson1992infectious}. An additional, often neglected, aspect of disease dynamics is the behavioral response of the host following the onset of illness. \emph{Sickness behaviors} such as fever, lethargy, depression, and loss of appetite, cause the host to limit their movement and social interactions \cite{doi:10.1093/aje/kwt196,HART1988123,AUBERT19991029,lopes2014socially}. These responses are thought to be adaptive as they protect host resources, reduce pathogen reproduction, and promote inclusive fitness by protecting the social group from infection\cite{10.1371/journal.pbio.1002276}.

\begin{figure*}[t]
\begin{center}
		\includegraphics[width=0.8\textwidth]{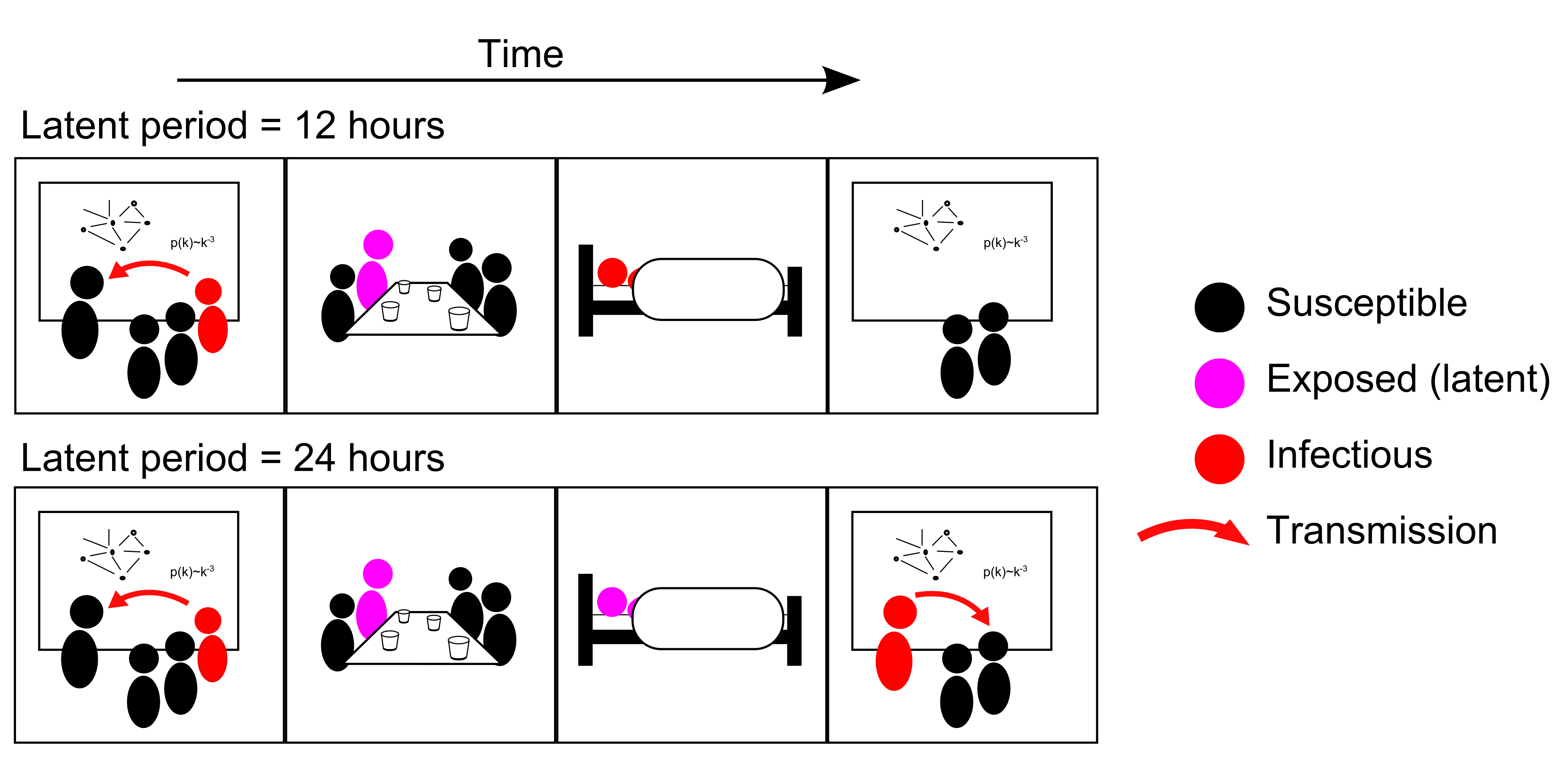}
        \caption{Conceptual illustration showing the effect of different latent periods. In the upper panel, the host becomes infectious $12$ hours after receiving the infection, at which point he has entered a more sedentary phase of his daily schedule. The symptoms of the infection influence him to avoid returning to his school or workplace and no further transmission occurs. In the lower panel, the infectious period begins at the same time of the day that he received the infection. While the symptoms of the disease may result in social withdrawal, there is a period of time for which he is both infectious and socially active, giving the disease an opportunity to spread.}
 \label{example}
\end{center}
 \end{figure*}

The analysis presented here considers the effect of social withdrawal behaviors on the spread of disease. While previous studies have considered reduced or rewired contacts \cite{doi:10.1080/17513758.2010.503376,scarpino2016effect}, we choose to examine the possibility that the infectious period, which is typically estimated from survey data or controlled experiments \cite{doi:10.1001/archinte.162.16.1842,carrat2008time,harris1996incubation}, is effectively cut short by the onset of sickness behaviors. Under these conditions it may be possible for diseases to have a reproductive advantage when their latent period synchronizes with the timing of human social behavior (see Figure \ref{example}). 

To test this hypothesis we use temporally resolved contact network data from various social settings; a conference, a school, and a hospital; and ask how disease spread would fare in each of these networks. By varying the length of the latent period in the model, we are able to observe a range of outcomes, with maximum disease risk for $24$-hour or $7$-day latent periods. 

\section{Methods}
\label{methods}
The goal of our study is to use empirical contact data to reveal the presence and magnitude of the synchronization effect. While the underlying concept may be relevant to a range of infectious diseases, we here focus specifically on lowly-transmissible respiratory disease. Data relating to the kind of close-proximity interactions that allow such diseases to transmit is publicly available (see Section \ref{data}). The data has the format of a temporal network, meaning that for each interaction recorded, we are given the identities of the two individuals involved, and the times for which the interaction started and ended.

We start by introducing a method to measure the contagion potential of a temporal contact network with respect to a disease with given latent and infectious period durations. We then introduce a method to quantify the effect of synchronization for simulated disease spread on the temporal contact network.

\subsection{Measuring disease impact through reachability}
\label{reachability_description}
The \emph{reachability} of an individual $X$ is defined as the number of individuals that could potentially become infected given that $X$ is the source of infection \cite{holme2005network,nicosia2012components,holme2012temporal,moody2002importance}. The original definition applied only to contagion processes where each individual can be in one of only two states: susceptible or infectious. The concept has since been developed to incorporate the recovered state \cite{koher2016infections}. The work presented here further extends the concept of reachability to include the latent state of the infection.

We compute the reachability of $X$ with respect to a given set of disease parameters, namely $\Delta_{E}$, which is the length of time for which the infection remains in the latent state after transmission (more commonly called the ``Exposed'' state), and $\Delta_{I}$, which is the length of time a host remains in the infectious state after the latent period has elapsed. 

We say that one member of the population, $X$, can ``reach'' another, $Z$, if a sequence of contacts exists that makes it feasible for the pathogen to have spread from $X$ to $Z$. For this to be the case, the time between any two consecutive contacts in the sequence must be greater than $\Delta_{E}$ and less than $\Delta_{E}+\Delta_{I}$ (see figure \ref{reachability_diagram}). We impose one further restriction that the sequence must begin during the period of length $\Delta_{E}$ that starts when $X$ is first observed interacting. This represents a situation in which $X$ arrives in the system in an infectious state.

We use an infectious period of $\Delta_{I}=2$. The host may experience illness for longer, but due to lack of social interactions they are effectively incapable of infecting others after this period. While this value has been selected arbitrarily, we consider it to be a reasonable amount of time for an individual to be infectious before sickness behaviors cause them to withdraw from social activity.

We have chosen to use reachability as it incorporates the relevant elements of disease dynamics while also being relatively fast to compute. It does, however, make implicit assumptions about the dynamics of the disease in question. Specifically, it is assumed that the disease variables, i.e. the latent and infectious period durations, are homogeneous across the population. To test whether the results obtained are sensitive to relaxations of these assumptions we have chosen to perform disease simulations on the data.

\subsection{Measuring the effect of synchronization}
\label{Simulation_description}
We use simulations of disease spread over the empirical contact network data to measure the magnitude of the effect of synchronization. Real diseases exhibit a range of dynamical behaviors that may invalidate the results obtained through reachability such as variance in the distribution of latent periods \cite{blythe1988distributed,lloyd2001realistic,10.1371/journal.pmed.0020174}, variance in the perseverance of individuals \cite{doi:10.1093/aje/kwt196}, and the proportion of individuals that are asymptomatic \cite{leung2015review,jacobs2013human}, denoted by $\sigma_{E}^{2}$, $1/\alpha_{I}$, and $a$, respectively. How each of these variables is incorporated into the simulation is described in Section \ref{disease_parameters}. 

At the beginning of the simulation, one individual is selected as the ``seed'' of infection and their infectious period begins at the moment of their first contact. During one second of contact between a susceptible individual and an infected host, the susceptible one becomes infected with probability $\beta$. Details of how  $\beta$ is chosen are presented in Section \ref{disease_parameters}. Once infection occurs the individual enters a latent period in which they are infected but unable to infect others, followed by an infectious period in which they can transmit to susceptible individuals they come into contact with. After the infectious period they enter the removed state. 

For each data-set, the list of individuals was ordered by name and the first $100$ were selected. For each individual $i$, we simulated the spread of disease originating from $i$ $100$ times with $\hat{\Delta}_{E}=10$ hours and another $100$ times with $\hat{\Delta}_{E}=22$ hours. To measure the magnitude of the outbreak we count the number of people who received an infection at any point in time with the exception of the seed and those who were infected directly by the seed. This choice has been made to exclude infections that have no relation to the latent period. The \emph{effect of synchronization} for an individual $i$ is defined as the mean outbreak magnitude when $\hat{\Delta}_{E}=10$ subtracted from the mean number of infections when $\hat{\Delta}_{E}=22$.

\subsection{Data}
\label{data}
\begin{figure*}[t!]
\centering
    \includegraphics[width=\textwidth]{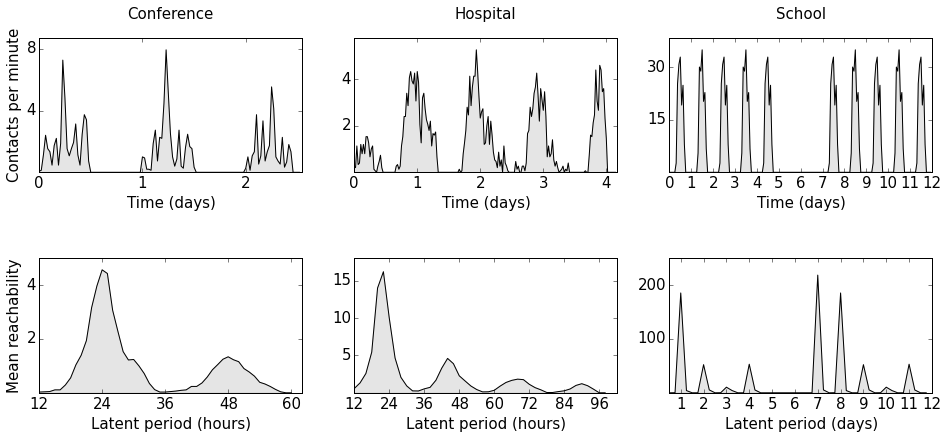}
   \caption{The top panels show the number of face-to-face interactions between pairs of individuals (in the school data we only show 2 of the 6 weeks). The bottom panel shows the mean reachability of a disease over a range of latent periods. For the purpose of presentation we have subtracted the number of nodes reached directly from the seed (this is the same for all values of the latent period duration). The tendency for latent periods which are larger multiples of 24 hours or 7 days to result in lower reachability is explained by the limited time span of the data-sets.}\label{reachability_fig}
\end{figure*}

We use human contact data from the Sociopatterns project (sociopatterns.org) recorded in three separate locations. Participants wore radiofrequency identification sensors that detect face-to-face proximity of other participants within 1-1.5 meters in 20-second intervals. Each data-set lists the identities of the people in contact, as well as the 20-second interval of detection. To exclude contacts detected while participants momentarily walked past one another, only contacts that are detected in at least two consecutive intervals are considered interactions. Sensors were not worn outside the locations being studied so there are long spans of inactivity corresponding to the periods from early evening to early morning (See the top panels of Figure \ref{reachability_fig}. 

Three data-sets were used: a conference in which $110$ participants were recorded over $3$ days \cite{isella2011s}, a hospital ward in which $74$ participants were recorded over $4$ days \cite{10.1371/journal.pone.0073970}, and a primary school in which $242$ participants were recorded over $2$ days \cite{Gemmetto2014,10.1371/journal.pone.0023176}. In a similar fashion to the original study \cite{Gemmetto2014}, we looped this data to to analyze the effect of latent period on a longer time scale. We used the first day to represent Monday, Wednesday and Friday and the second day to represent Tuesday and Thursday. We then added 2 days of inactivity to replicate a typical school week and weekend. We repeated this week to form 6 weeks of data. 

\begin{figure*}[t!]
\centering
	\includegraphics[width=\textwidth]{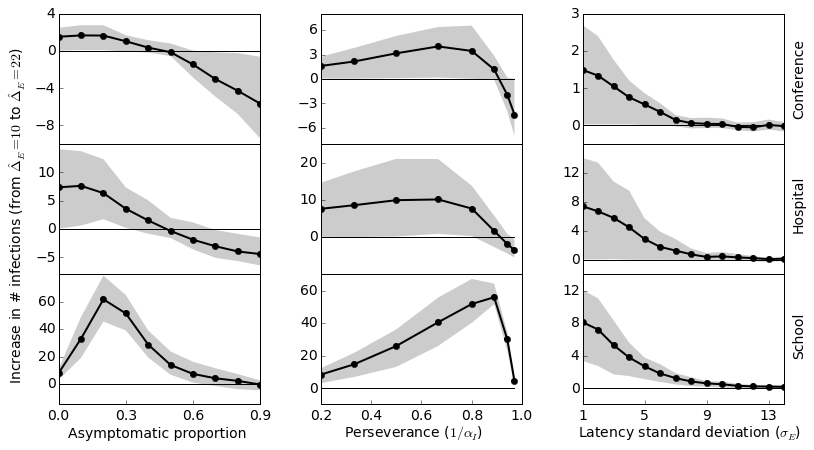}	
    \caption{The effect of synchronization. The dark line represents the mean effect size over individuals in the population. The effect size is defined as the increase in mean outbreak size between a disease for which the latent periods are gamma distributed with a mode of $10$ hours, and one which has a mode of $22$ hours (variances are equal). Points correspond to the values on the horizontal axis for which the effect size was computed and the gray area is the inter-quartile range. In general, we see that the synchronization effect observed in Section \ref{reachability_results} is present for a wide range of parameters values in the disease model. \label{simulation_plot}}
\end{figure*}

\section{Results}
\label{results}
In all $3$ empirically observed contact networks, the rate of contact between individuals fluctuates periodically in time with the cycles of human social activity (see Figure \ref{reachability_fig}). By considering the spread of disease through these networks, we find that the impact of infectious disease is maximized when timing of infectiousness is synchronized with these temporal dynamics. We show this through analytical measurement of the temporal network structure (reachability) and test the robustness of these results through analysis of simulated disease outbreaks on each network.

\subsection{Influence of the latent period on disease impact}
\label{reachability_results}
To reveal the epidemiological impact of the timing of infectiousness, the mean reachability (over all individuals in the network) was computed for a range of latent periods. The results are shown in Figure \ref{reachability_fig}.

For the conference setting, we observe two maxima. The peak corresponding to a latent period of just under $24$ hours is larger than the size of the peak corresponding to an approximate $48$-hour latent period. In the first case, those infected by the seed on day one, can cause a second generation of infections on day two, which then causes a third generation infections on day three. In the second case only two generations of infection are possible.

Similarly, in the hospital setting we observe $4$ maxima. The largest peak corresponds to a 24-hour latent period and represents generations of infection that reproduced on days $2$-$5$ of the hospital contact data-set. The peak corresponding with a $48$-hour latent period represents a disease that was able to reproduce on only $2$ of the $5$ days of the hospital contact data-set ($2$ and $4$ days after the time the seed was infectious). The final two peaks represent the optimal latent periods in cases where only $1$ generation of infections is possible due to the limited time frame of the data.  

Results for the school setting show a comparable pattern but on a larger time-scale. In this data all individuals are seeded on day $1$; the peak corresponding with a 24-hour latent period includes infections that occurred on days $2$-$5$; since no contacts occur on weekends, the disease dies out on Day $6$ (Saturday). Peaks corresponding to latent periods of $2$ days and $4$ days are only able to reproduce on $3$ days of infectious contacts. In both scenarios, the disease also dies out over the weekend. 

The absolute maximum reachability corresponds with a latent period of $7$ days. This peak occurs because there are effectively $6$ days ($6$ Mondays) of infectious contacts during which the disease can reproduce. Individuals infected during periods of high social activity on the first Monday infect individuals during periods of high social activity on the subsequent Monday, and the process repeats. 

\subsection{Robustness of the observed synchronization}
We computed the effect of synchronization for simulated diseases over a range of parameter values. For the initial baseline parameters; the mode of the infectious duration distribution $\hat{\Delta}_{I}=2$ hours, standard deviation of the latency distribution $\sigma_{E}=1$ hour, asymptomatic proportion $a=0$, perseverance $1/\alpha_{I}=0.2$; the effect of synchronization was positive, i.e. changing the mode value of the latent duration distribution from $\hat{\Delta}_{E}=10$ to $\hat{\Delta}_{E}=22$ yields an increase in the number of infections. This is consistent with the results of section \ref{reachability_results}. We then tested the robustness of this effect by perturbing each parameter away from its baseline value towards more realistic scenarios. Figure \ref{simulation_plot} shows that the effect of synchronization remains present under most of the parameterizations we tested.

Since asymptomatic individuals remain in the system for significantly longer than those who do experience symptoms, their presence increases the overall outbreak size. Moreover, since the time at which they first become infectious has relatively little affect on the number of infections they cause, we expect to see a decrease in the effect of synchronization as $a$ increases. This was found to be the case beyond some value ($0.1<a<0.3$), however, at lower values the presence of asymptomatic individuals did not benefit the non-synchronized disease any more than the synchronized one. In the school setting, the synchronization effect was even amplified by the addition of asymptomatic cases.

Increasing perseverance causes individuals to remain in the system for a prolonged duration, and consequently allows more opportunities for the infection to spread. In Figure \ref{simulation_plot}, the relationship between perseverance and synchronization effect is qualitatively similar to that seen for the asymptotic proportion; at low values, perseverance benefits both the non-synchronized and synchronized diseases. For larger values, the individuals who persevere for prolonged durations dominate transmission and the time at which they first become infectious loses its significance.

Variance in the latent period, $\sigma_{E}$, affects the mean outbreak size differently depending on the value of $\hat{\Delta}_{E}$. For $\hat{\Delta}_{E}=10$, as the variance increases the probability that the latent period will be close to $10$ gets smaller and the probability that the infectious period of individual will intersect with a period of high social activity gets larger. The opposite is true for $\hat{\Delta}_{E}=22$. Increasing $\sigma_{E}$ therefore causes the mean outbreak sizes, of the two cases, to converge towards each other and the synchronization effect to converge to zero.

In the conference and hospital settings, the number of generations of infection is limited by the duration of the data (3 and 4 days, respectively). In general, this gives the shorter latent period an advantage over the longer one and we eventually see the synchronization effect go below zero. In the primary school dataset, on the other hand, most infections die out before the end of the 6-week duration and the effect of synchronization is rarely negative. 

\section{Discussion}
\label{discussion}
Our results support the hypothesis that disease risk can be amplified by synchronization between the latent period of the infection and the circadian rhythms of the host population. This conclusion has implications regarding the way infectious disease should be modeled at the population scale; contact rates vary dynamically, both periodically according to cycles of human behavior, and in response to the disease itself. Consequently, the window of opportunity for transmission may be much shorter than the actual duration of infectiousness determined in experimental studies; thus, the important question is not ``how long is the infectious period?'', but ``when does the infectious period begin?''. 

Epidemic models that do not consider the effect of sickness behaviors, and assume a constant (non cyclic) rate of contact (for example, \cite{ferguson2005strategies}), may be at risk of identifying the wrong primary drivers of transmission. This is most likely to apply to cases where sickness behaviors are strong and infectiousness begins shortly before the onset of symptoms. It has been shown in some diseases that there is a period of time between beginning of infectiousness and the onset of symptoms \cite{lau2010viral,harris1996incubation} and others have demonstrated the epidemiological significance of this of period \cite{Fraser20042004}. The present work suggests that, in such cases, the duration of the latent period is a primary driver of infection.

Since the present work is limited by the incompleteness of the data, the lack of empirical disease data, and a lack of knowledge about illness induced behavior change, we suggest that this hypothesis should be scrutinized in any way possible. In particular, the data we have used does not include interactions that occur outside of the established setting; the effect of these missing links could be significant enough to break the synchronization pattern we see. For example, interactions that occur in the household may act as bridges that allow the infection to spread between schools and workplaces \cite{Pellisrsif.2008.0493}. This increase in mobility may contribute more to the size of the outbreak than does the effect of synchronization. 

If, on the other hand, synchronization does give a disease a reproductive advantage, then we would expect many epidemic diseases to have adapted to have a latent period that aligns with the timing of circadian cycles. We compiled information about the latent period of a number of infectious diseases (Table \ref{disease_table} and Figure \ref{multiplot}) and no clear trend is apparent, however, we were unable to draw a firm conclusion based on such a small amount of information.

Finally, we suggest the possibility that control strategies for managing infectious disease may be designed around the effects of synchronization. By manipulating the schedule of people in a school or workplace in a way that breaks the synchronization pattern, it may be possible to create an environment that is hostile towards infections with particular latent periods. More generally, a better understanding of the coupling between human and disease dynamics could lead to methods of social distancing that are sensitive to the temporal dynamics of infectious disease.

\subsection*{Declarations}
\small
\subsubsection*{Author contributions}
KS conducted the analysis of reachability and drafted the manuscript. EC created and analyzed the disease simulations. EC and SB conceived and designed the analysis. KS and EC  contributed equally to the writing of the manuscript. 
\subsubsection*{Funding}
EC received funding through NSF Grant No. 1414296 as part of the joint NSF-NIH-USDA Ecology and Evolution of Infectious Diseases program.
\end{multicols}
\small
\bibliography{bibfile}
\bibliographystyle{ieeetr}

\newpage
\clearpage
\pagenumbering{arabic}
\renewcommand*{\thepage}{S\arabic{page}}
\renewcommand\thefigure{S\arabic{figure}}
\renewcommand\thetable{S\arabic{table}}
\renewcommand\thesection{S\arabic{section}}

\setcounter{figure}{0}  

\begin{multicols}{2}
\section*{\huge Supplementary information}
\setcounter{section}{0} 
\section{Model parameters for simulation}
\label{disease_parameters}
\subsection{Transmissibility}
The value of $\beta$ is calculated for each data set separately in such a way that $R_{0}$, the expectation of the number of secondary infections caused by one infected individual, is equal to $1$. This is calculated by solving
\begin{equation}
\frac{\text{Total interaction duration}}{\text{Population size}\times \text{Total duration}}\times\langle \Delta_{I}\rangle\times\beta=1
\end{equation}
The quotient on the left hand side is the probability that an individual, selected randomly at a random point in time, will be engaged in contact. This is then multiplied by the amount of time they would be infectious should they receive the infection, and again by the probability that transmission will occur during one second. This gives $\beta=0.0049$ for the conference data, $\beta=0.0019$ for the hospital, and $\beta=0.0009$ for the primary school.

Note that the calculation of $R_{0}$ averages over the entire duration of the data, including periods of both high and low activity. In practice if host is infectious during a high activity period then the the number of secondary infection we would expect them to cause will be significantly higher than $1$. Conversely, if are infectious during a time of low activity then it will be significantly lower.

\subsection{Asymptomatic proportion}
Some members of the population may show no signs of infection (up to $28\%$ reported for influenza \cite{leung2015review} and $32\%$ for rhinovirus \cite{jacobs2013human}), or might just ignore them completely, in which case their behavior does not change. At the beginning of the simulation, a random sample of the population are chosen to be asymptomatic. These individuals, who make up a fraction $a$ of the total population, have an infectious period of $5$ days.

\subsection{Latent period duration variance}
The duration of the latent period may vary between individuals depending on their age, gender, or other characteristics \cite{blythe1988distributed,lloyd2001destabilization,lloyd2001realistic,10.1371/journal.pmed.0020174}. In the simulation, the latent duration for each infected individual is drawn from a gamma distribution with mode $\hat{\Delta}_{E}$ and standard deviation $\sigma_{E}$. In general, the gamma distribution gives non-zero probability of selecting values close to $0$ and, as a consequence, large values of $\hat{\Delta}_{E}$ are only possible when $\sigma_{E}$ is also sufficiently large. To achieve full control over both parameters, we modify the selection procedure for cases where $\hat{\Delta}_{E}>5\sigma_{E}$ by decomposing $\hat{\Delta}_{E}$ into the sum of a non-random component of length $\hat{\Delta}_{E}-5\sigma_{E}$ and a gamma distributed random variable with mode $5\sigma_{E}$.

\subsection{Perseverance}
Once infected, the behavioral response of individuals may vary; some might leave the system (or take other measures to prevent infection) immediately, whereas some may remain a risk to others for a more prolonged duration \cite{doi:10.1093/aje/kwt196}. In the simulation, the duration of the infectious period of each individual is randomly selected from a gamma distribution with a mode of $\hat{\Delta}_{I}=2$ hours. We define perseverance as $1/\alpha_{I}$ where $\alpha_{I}$ is the shape parameter of the gamma distribution. While the mode does not change, increasing the perseverance fattens the the tail of the distribution. Consequently the mean and variance of the distribution also increase. 
\end{multicols}
\begin{figure}[H]
		\includegraphics[width=\textwidth]{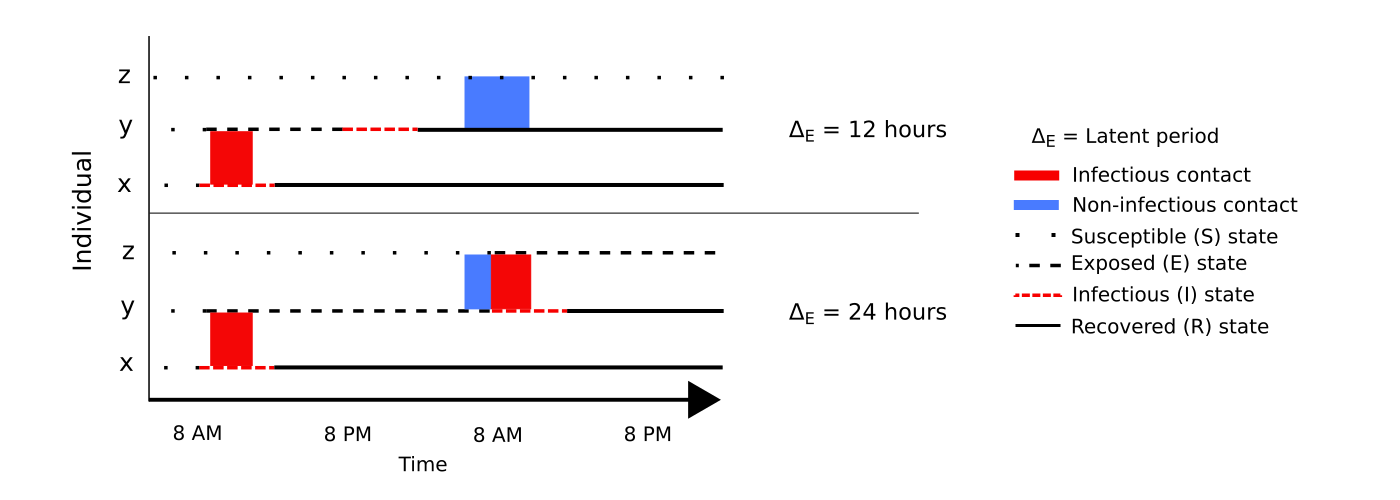}
        \caption{An example showing the difference between a 12 and 24 hour latent period. For both scenarios, we consider the same contact sequence. The color and pattern of the line represent the disease state of the individual. The color of the shaded region between two lines represents an either a  potentially interaction (in the bottom panel the infectious period of Y begins during the interaction between Y and Z). Individual X is infected on the morning of the first day. When infected with a disease with a 24-hour latent period, the infection can reach both individuals Y and Z; however, with a 12-hour latent period, it can only reach individual Y.}
 \label{reachability_diagram}
 \end{figure}

\begin{figure}[t]
\centering\includegraphics[width=0.95\textwidth]{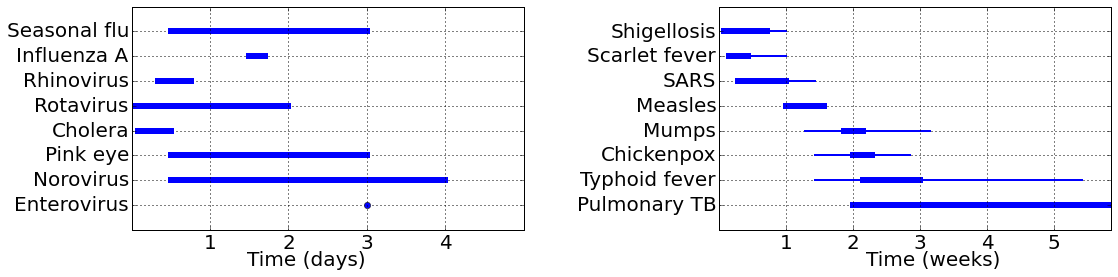}	
    \caption{Latent periods of infectious diseases based on information gathered from the literature. The bold line represents the longest possible latent or infectious period that can be estimated with high certainty. The thin line represents that the latent or infectious period could be longer but there is disagreement among sources. A dot signifies that a single length of time, rather than a range, was provided. Time is in days for the top panel and in weeks for the bottom. The latent period range for pulmonary TB was shortened for ease of viewing (actual upper bound is $12$ weeks). All information shown, as well as incubation and infectious periods, are provided in Table \ref{disease_table}.
 \label{multiplot}}
 \end{figure}

\newpage
\begin{table}[b]
        \caption{The maximum range of latent, incubation, and infectious periods, where available, for several transmissible diseases, as collected from the literature.\label{disease_table}}
    \begin{tabular}{ | p{4cm} | p{4cm} | p{4cm} | p{4cm} |}
    \toprule
    \textbf{Disease} & \textbf{Latent period} & \textbf{Infectious period} & \textbf{Incubation period}\\
		\midrule
    Rhinovirus  & 8 - 18 hours \cite{harris1996incubation} & 7 - 14 days \cite{jacobs2013human} & 1.4 - 2.4 days  \cite{lessler2009incubation} or 16 hours \cite{harris1996incubation} \\  
   Seasonal flu \cite{carrat2008time} & 0.5 - 3 days, average 1.1 days & 4.31 - 5.29 days, average 4.8 days & 1 - 3 days  \\
   Influenza A (H1N1 and H3N2) \cite{cori2012estimating} & 1.5 - 1.7 days, mean 1.6 days & 0.5 - 1.7 days, mean 1.0 day & not provided \\ 
   Pink eye (bacterial) \cite{WebMD, Mayo} & 24-72 hrs & 7 - 10 days or until 24 hours after start of antibiotics & 24-72 hrs \\
   Pink eye (viral) \cite{WebMD, Mayo} & 12 - 72 hour & usually 5 - 7 days, can become chronic & 12 - 72 hour    \\
   Enterovirus D68 \cite{CDC} & 3 days & 1 - 3 weeks, average 10 days & 3 - 6 days, others say 3 - 10 days   \\
    Rotavirus \cite{CDC,WebMD} & 48 hours or less & 6 - 11 days & 48 hours \\
     Scarlet fever \cite{WebMD} & 12 hours - 7 days & 1 - 2 weeks & 12 hours - 7 days, usually 2 - 5 days \\
    Cholera \cite{CDC, WebMD} & 2 hours & 7 - 14 days & 2 hours – 5 days, usually 2 - 3 days\\
    Norovirus \cite{CDC} & 12 - 94 hours &  4 - 6 days & 12 - 48 hours\\
   Shigellosis \cite{CDC} & usually 1 - 3 days, up to 7 days & 2 days up to ``a few weeks'' & usually 1 - 3 days, up to 7 days\\ 
    SARS \cite{CDC} & usually 2 - 7 days, up to 10 days & 2 - 21 days & usually 2 - 7 days, up to 10 days\\ 
    Measles \cite{CDC} & 6 - 15 days & 9 days & 7 - 14 days (cold-like symptoms), 10 - 19 days (rash)\\  
    Chickenpox \cite{WebMD} & 10 - 21 days & 6 - 8 days & 10 - 21 days\\ 
    Mumps \cite{CDC} & 9 - 22 days, usually 13 - 15 days & 8 days & 12 - 25 days, usually 16 - 18 days\\
    Typhoid fever \cite{CDC} &  10 days to about 1 week after symptoms onset & usually 2 weeks, can be up to 1 year & usually 15 - 21 days  \\
    P. tuberculosis \cite{Mayo, CDC} & 2 - 12 weeks &  2 weeks  & 2 - 12 weeks\\
		\bottomrule
     
    \end{tabular}
\end{table}

\end{document}